\documentclass[10pt]{article}
\usepackage{fullpage}
\usepackage{amsmath}
\usepackage{amssymb}
\usepackage[dvips]{epsfig}
\usepackage{color}

\def\##1{\underline #1}
\def\=#1{\underline{\underline #1}}

\def\.{\mbox{ \tiny{$^\bullet$} }}

\begin{document}
\begin{center}

{\bf {\LARGE Phase diagram of magnetic configurations for soft magnetic nanodots of circular and elliptical shape obtained by micromagnetic simulation}}

\vspace{10mm} \large

 E.R.P. Novais\footnote{E--mail: erpn@cbpf.br} and A.P. Guimar{\~a}es\footnote{E--mail: apguima@cbpf.br}\\
{\em Centro Brasileiro de Pesquisas F{\'{\i}}sicas,\\  R. Xavier Sigaud 150,  22290-180, Rio de Janeiro, Brazil.}\\

\end{center}

\vspace{4mm}

\normalsize
\begin{abstract}
Magnetic disks or dots of soft magnetic material of sub-micron dimensions may have as the lowest energy magnetic configuration a single-domain structure, with magnetization either perpendicular of parallel to the plane, or else may form magnetic vortices. The properties of these vortices may be used to encode data bits, in magnetic memory applications. In the present work the OOMMF code was used to compute by micromagnetic simulation the energy and the magnetization of circular and elliptical nanodots of permalloy.  For the elliptical magnetic dots the analysis was made for variable thickness and length of major axis, keeping a 2:1 axis ratio.  From the simulations, a phase diagram was constructed, where the ground state configurations of the nanodots are represented in a diagram of nanodot height versus length of the major axis $2a$ of the ellipse. The phase diagram obtained includes regions with one and two vortices; it is similar, but more complex than that derived using a numerical scaling approach, since it includes configurations with lateral vortices. These diagrams are useful as guides for the choice of dimensions of elliptical nanodots for practical applications.

\end{abstract}

% --------------------------------------------------------------%
\section{Introduction}
% --------------------------------------------------------------%
Nanoscopic and mesoscopic magnetic structures in the form of dots of circular or elliptical shape have attracted the interest of many workers in recent years in view of their very interesting physical properties and for their potential applications; for an introduction see \cite{Chien:2007}. Quasi-twodimensional magnetic objects of nanoscopic or sub-micron dimensions (referred here as nanodots) made of soft magnetic materials, such as permalloy, may present, for their lowest energy configuration, a single-domain ordering, with magnetic moments either perpendicular or parallel to the plane of the dot. For single-domain, we understand the magnetic structure where the spins are in general parallel, although deviations of the spins near the surfaces are expected. They may also, depending on the dimensions, form magnetic vortices, or swirls.

These magnetic vortices appear as lowest energy magnetization configurations and have been observed by many experimental techniques, such as magnetic force microscopy \cite{Shinjo:2000}, X-ray microscopy \cite{Choe:2004}, or inferred from hysteresis curves \cite{Cowburn:1999}. The magnetic vortices have a core that is magnetized perpendicularly to the plane of the nanodot, with a radius of the order of the exchange length in the thin dot limit \cite{Hubert:1999}, about 5 nm in the case of permalloy. Applied magnetic fields or spin polarized currents may reverse the direction of magnetization of this core; otherwise, the magnetic vortices are stable structures.

The proposed applications of magnetic nanodots include their use in patterned magnetic recording media \cite{Thomson:2008} and as elements in magneto-resistance random access memories (MRAM's) \cite{Kim:2008}, \cite{Bohlens:2008}. Nanodots of circular shape  present flux closure, and this implies that the perturbation from neighbor elements is reduced, allowing the use of higher density arrays, an important consideration in applications.

Nanodots exhibit different magnetic structures, depending on their dimensions and geometry. Consider, for example, 20 nm-thick circular nanodots prepared from a soft magnetic material, such as permalloy. Below a radius of 10 nm, the magnetization is perpendicular to the plane, from 10 to about 30 nm the magnetization is parallel to the plane, and above this radius, the minimum energy magnetic structure is a vortex.

Magnetic vortices are structures characterized by a direction of rotation (circulation $c$), that may be clockwise (CW, $c=-1$) or counterclockwise (CCW, $c=+1$), the direction of the magnetization of the vortex core, that may be along the positive $z$ axis or antiparallel to it; this defines the polarity as $p=+1$ or $p=-1$, respectively. The core has a radius of the order of the exchange length of the material $l_{\mathrm ex}=\sqrt{2A/\mu_0 M_s^2}$ (or $l_{\mathrm ex}=\sqrt{2A/4 \pi M_s^2}$ in the CGS), where $A$ is the exchange stiffness constant and $M_s$ is the saturation magnetization. With the parameters used for permalloy in the present work, $l_{\mathrm{ex}}=5.29$ nm. The combination of circulation and polarity defines the chirality, or handedness of the vortex; a vortex may be righthanded ($cp=+1$) or lefthanded ($cp=-1$); one should note that the word chirality is sometimes used in the literature to describe the sense of rotation of the vortex.

The different magnetic configurations, and consequently the large variation in the magnetic properties observed in nanodots as a function of dimensions, underlines the interest in the study of diagrams (phase diagrams) mapping the parameter space where a given magnetic behavior is to be expected. We have chosen in the present work to draw such diagrams for nanodots of circular and elliptical shape using micromagnetic simulation. This simulation allows the computation of the total energy of the nanodot, which encompasses a magnetostatic contribution, the exchange contribution and the anisotropy term. Micromagnetic simulations are invaluable for the prediction of magnetic behavior of these nanostructures, although necessarily containing some idealizations, for example, assuming the constancy of the thickness of the dots, homogenous physical properties, or the absence of defects.

In the present work the energy per unit volume and the equilibrium magnetic structures of thin permalloy islands of circular or elliptical shape and sub-micron dimensions were studied using the micromagnetic simulation program OOMMF, a free software available from NIST \\cite{oommf}. The crystalline anisotropy term in the energy was neglected, as it is usual in simulations of permalloy samples.

% --------------------------------------------------------------%
\section{Results and discussion}
% --------------------------------------------------------------%
We have made a systematic study of the magnetic configuration of circular and elliptical (of axial ratio 2:1) nanodots of permalloy through micromagnetic simulations.
For the simulations we used the 3D OOMMF code. The computation was performed for permalloy samples, using the exchange stiffness constant $A=1.3\times10^{-11}$ J/m, the
saturation magnetization $M_s=860\times 10^3$ A/m and the anisotropy constant $K=0$, the typical values used in the program. To make the present results of more general use, they have been given in terms of parameters normalized using the exchange length $l_{\mathrm ex}$.  The cell size in most cases was taken as $5\times 5 \times 5$ nm$^3$; the dependence of the simulation results on the cell size was verified for some configurations.

For some dimensions of the nanodots, the simulation may converge to a configuration that does not correspond to the absolute energy minimum. Therefore, the simulations made with parameters near the boundary regions between different configurations had to be made by first imposing every possible configuration, and afterwards comparing the resulting energies to determine the spin arrangement corresponding to the absolute energy minimum.

To save computation time, the simulation for the larger elliptical dots used as a starting configuration a multidomain structure that was close to a single or double vortex structure, depending on the value of the major axis of the ellipse whose magnetic structure was being simulated.  Some checks with the full calculation were made to verify that this procedure did not lead to secondary energy minima.

% --------------------------------------------------------------%
\subsection{Circular nanodots}
% --------------------------------------------------------------%
Based on the shape of the hysteresis curves of nanodisks of Ni, Co, CoNi and CoP of different thicknesses and diameters, Ross et al. \cite{Ross:2002} have drawn a magnetic configuration phase diagram. The diagram contained three configurations, described as in-plane flower, out of plane flower, and vortex/multidomain. The boundaries between the regions corresponding to these arrangements agreed with the frontiers computed by micromagnetic simulation.

Metlov and Guslienko \cite{Metlov:2002} derived an expression for the total energy of circular nanodots, and from this result obtained a phase diagram with regions of in-plane magnetization, perpendicular magnetization and vortex structure. Scholz et al. \cite{Scholz:2003} computed the equilibrium magnetic configuration for some dimensions of dots by micromagnetic simulation and found a general agreement with the phase diagram derived by Metlov and Guslienko \cite{Metlov:2002}.

Another simulation study, this time using a scaling approach, was made for circular and elliptical nanodots \cite{Zhang:2008}. For circular dots the phase diagram presented in that work contained parallel, perpendicular and vortex magnetization structures; their results for elliptical dots will be commented below.

\begin{figure}[!h]
\begin{center}
\includegraphics[width=11cm]{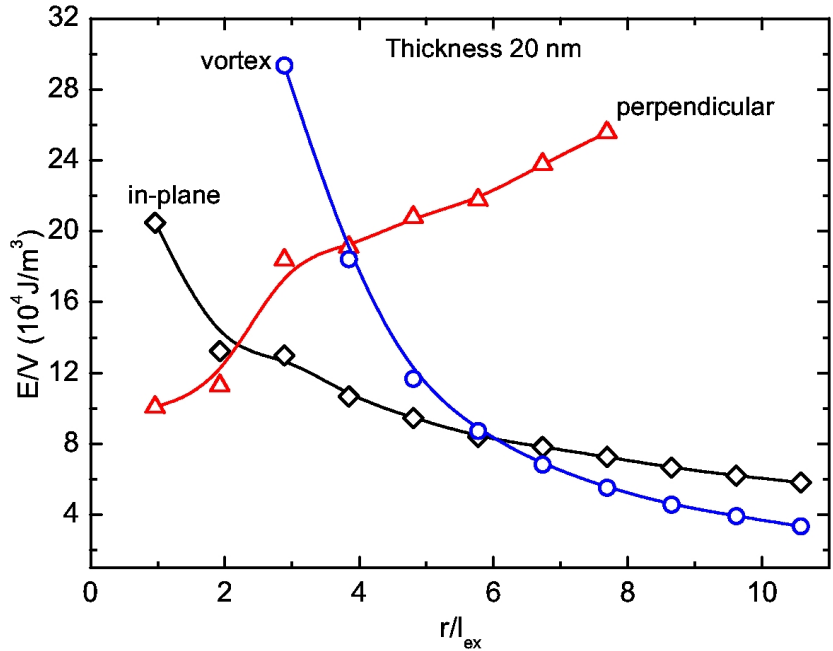}
\end{center}
\caption{(Color online) Energy per volume versus reduced radius ($r/l_{\mathrm ex}$) for 20 nm-thick circular nanodot of permalloy, computed
by micromagnetic simulation. The quantity $l_{\mathrm ex}$ is the exchange length.} \label{fig:EnergyVsRadius}
\end{figure}

\begin{figure}[!h]
\begin{center}
\includegraphics[width=11cm]{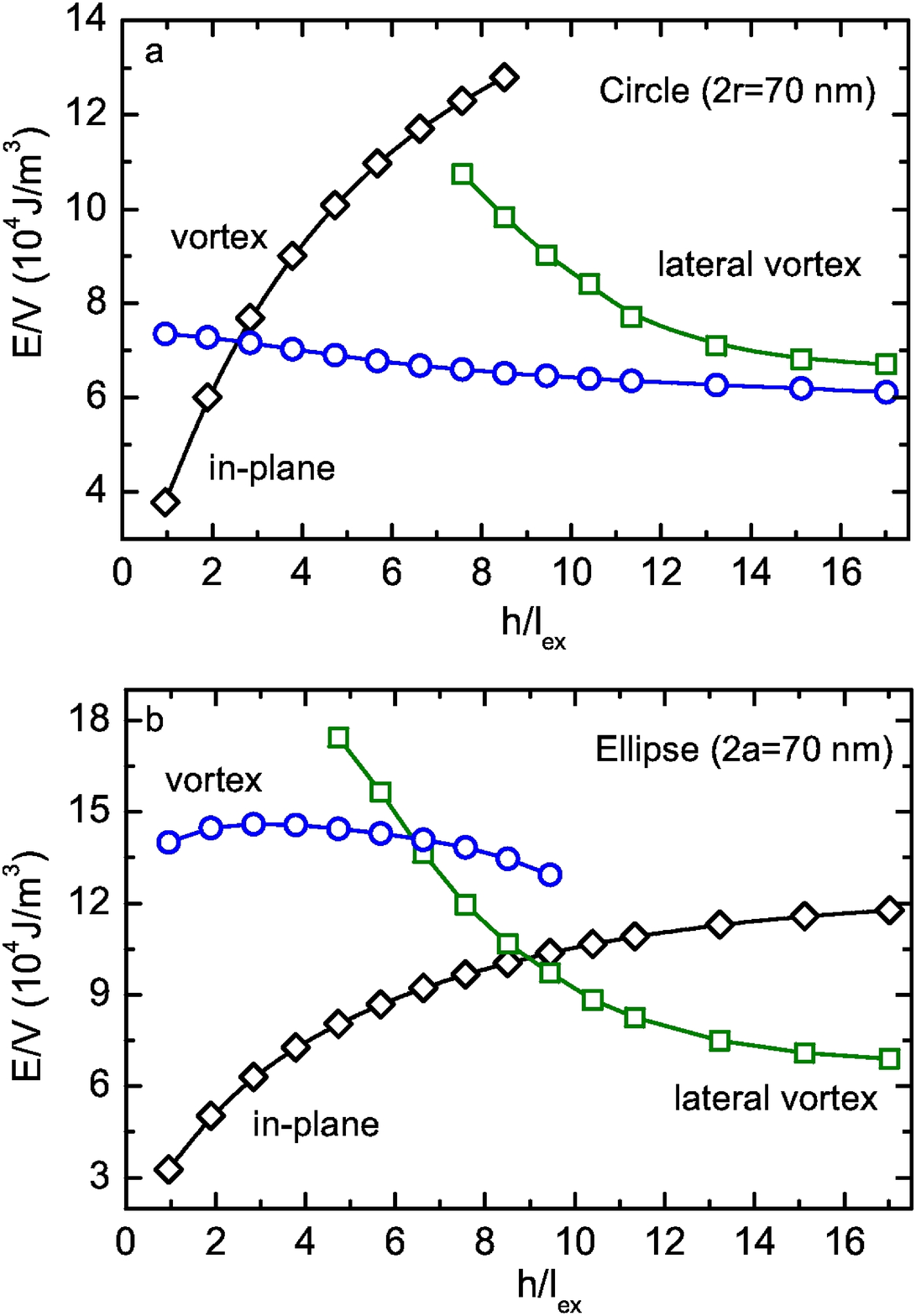}
\end{center}
\caption{(Color online) Energy per unit volume for circular and elliptical permalloy nanodots versus reduced height $h/l_{\mathrm ex}$ for a) 70 nm-diameter circular nanodot and b) elliptical nanodot of major axis $2a=70$ nm. The curves were computed with micromagnetic simulation.} \label{fig:EnergyVsThick}
\end{figure}

Our results for the total energy per unit volume  ($E/V$), or energy density, for some dimensions of circular permalloy nanodots are presented in Figs. \ref{fig:EnergyVsRadius} and \ref{fig:EnergyVsThick}. The first figure shows the variation of $E/V$ versus reduced radius $r/l_{\mathrm ex}$ for a 20 nm thick nanodot, for three magnetization configurations; it is evident that the energy of the minimum energy configuration changes only by a factor of about 3 in the range of radii shown. The same is valid for the second figure, where $E/V$ is drawn against reduced thickness $h/l_{\mathrm ex}$. The range of energy per unit volume for each given magnetic configuration is wider, up to about 10 times in some cases, for the dimensions shown in the figures.

The computed curves of energy density versus thickness, or height, are shown, for circular nanodots, in Fig. \ref{fig:EnergyVsThick}a; the height is measured in units of the exchange length $l_{\mathrm ex}$.

\begin{figure}[!h]
\begin{center}
\includegraphics[width=11cm]{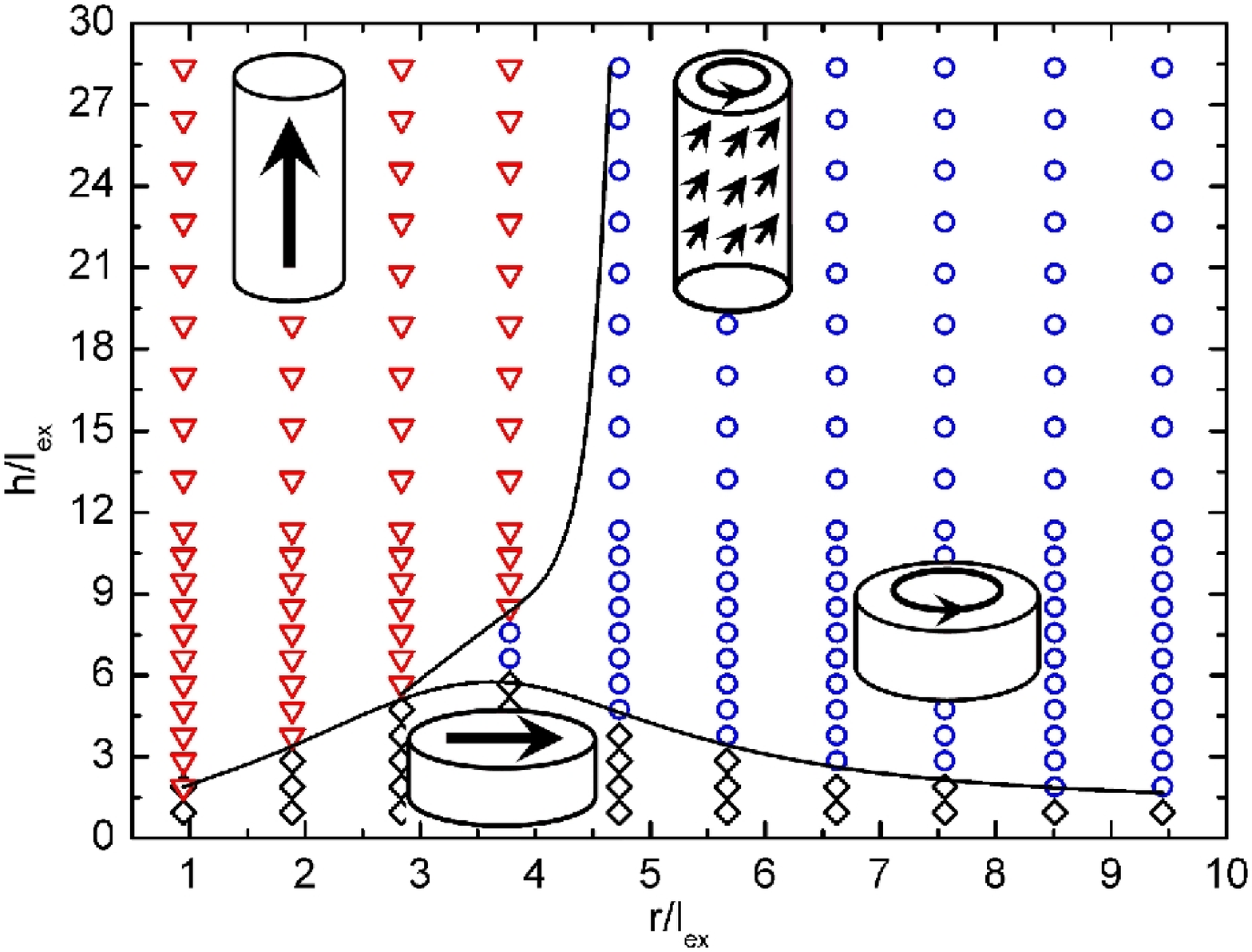}
\end{center}
\caption{(Color online) Phase diagram for circular nanodots of permalloy, as a function of reduced height $h/l_{\mathrm ex}$ and reduced radius $r/l_{\mathrm ex}$,
drawn from the minimum energy computed with the simulation. The continuous line was drawn following
a procedure described in the text.} \label{fig:DiskDiagram}
\end{figure}

Fig. \ref{fig:DiskDiagram} shows the phase diagram for circular magnetic nanodots, obtained from the computed energies for the different magnetization configurations. The diagram shows three regions, corresponding to in-plane single domain, out of plane single domain, and magnetic vortex.
The boundary lines between the several magnetic configurations exhibited in Fig. \ref{fig:DiskDiagram}  were obtained from the intersection of the curves of $E/V$ versus dimension for the different arrangements.

% --------------------------------------------------------------%
\subsection{Elliptical nanodots}
% --------------------------------------------------------------%
Experimental studies using an array of micrometric elliptical permalloy nanodots with 2:1 axis ratio showed both single vortex and double vortex states, the latter configuration being more probable when observed (at remanence) after saturation in the direction of the long axis \cite{Buchanan:2005}.

Magnetic configurations with one, two and three vortices were observed experimentally with the MFM technique on permalloy ellipses some microns long \cite{Lai:2007}. In the same study these magnetization arrangements were also derived from micromagnetic simulations performed with permalloy ellipses of 720x240x30 nm$^3$.
\begin{figure}[!h]
\begin{center}
\includegraphics[width=11cm]{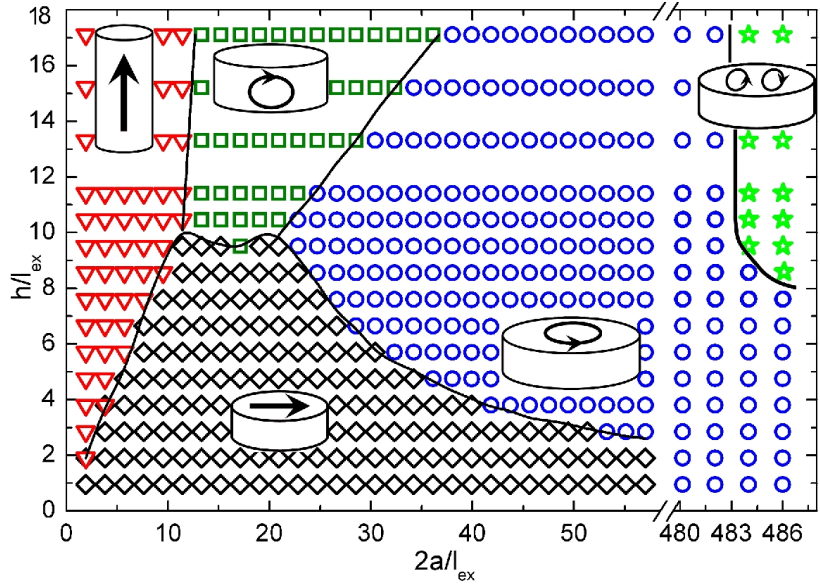}
\end{center}
\caption{(Color online) Phase diagram for elliptical nanodot of permalloy, as a function of the reduced height $h/l_{\mathrm ex}$ and reduced length
of the major axis of the ellipses $2a/l_{\mathrm ex}$. The ellipses in every case have axes in the ratio 2:1. Note the break in the horizontal scale, to reveal the appearance of two-vortex structures obtained for larger ellipses. The continuous line was drawn following a procedure described in the text.}
\label{fig:EllipseDiagram}
\end{figure}

Earlier work on elliptical dots using micromagnetic simulation sketched a phase diagram containing  three configurations: a region of quasi-single domain structure, single domain and vortex. The computation was made for fixed nanodot thickness (30 nm) \cite{Usov:2001}.

The phase diagram obtained by Zhang et al. \cite{Zhang:2008} for elliptical dots shows, besides the configurations observed for circular dots, a double vortex magnetic arrangement, which appears, for the parameters chosen in their illustration (Fig. 7, in \cite{Zhang:2008}) for major axes $2a$ above about 300 nm. In their simulation, the contribution of the vortex core to the energy was neglected; the authors estimated that the inclusion of the core contribution would displace the boundaries in the phase diagram by about 35 \%, quite a significant shift.

The computed curves of energy density versus thickness, or height, are shown, for elliptical nanodots, in Fig. \ref{fig:EnergyVsThick}b; the height is measured in units of the exchange length $l_{\mathrm ex}$. The comparison between the graph for circular and elliptical dots shows that the configuration of lateral vortex is a minimum energy configuration for elliptical dots, but not for circular dots. In the work of Ha et al. \cite{Ha:2003} using another computer code, in one single case a lateral vortex was observed in simulations for circular dots; however, their methodology was different, involving the determination of the magnetization configuration at remanence, after saturating the nanodot with an external magnetic field. This difference may explain also other differences in the phase diagram that they have obtained.

The phase diagram obtained from our micromagnetic simulations for elliptical nanodots is shown in Fig. \ref{fig:EllipseDiagram}; this diagram shows single-domain arrangements, both in-plane and perpendicular, as found in simulations for the circular dots. However, it has a more complex structure, with configurations with lateral vortices and, for the larger dots, configurations with more than one vortex; the region where two vortices appear is shown in the figure.

\begin{figure}[!h]
\begin{center}%
\includegraphics[width=11cm]{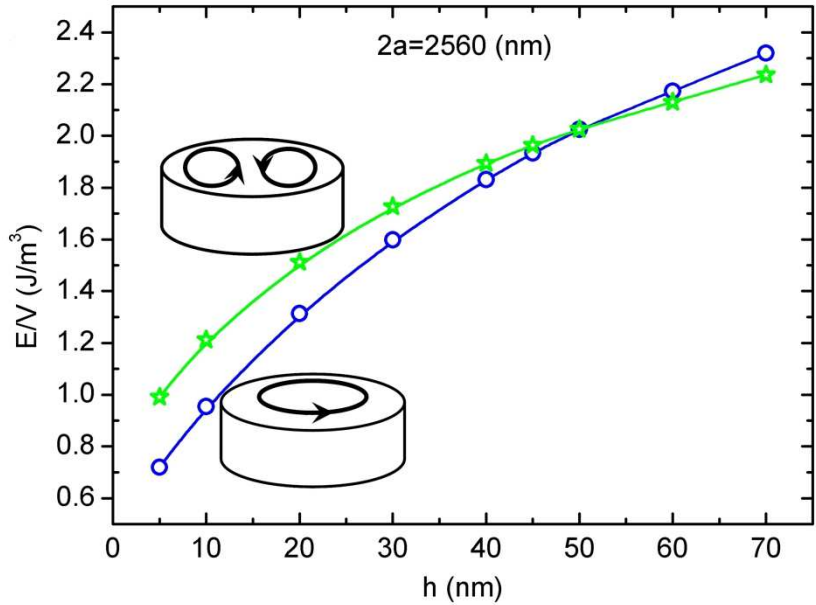}
\end{center}
\caption{(Color online) Energy per volume ($E/V$) versus thickness $h$ of elliptical nanodot, with axis $2a=2560$ nm, showing that above about $h=50$ nm, the configuration with two vortices has a smaller energy density than the configuration with one vortex, from the micromagnetic simulation.}\label{fig:onetwo}
\end{figure}

We have also shown, in the graph for elliptical nanodots, the curves of energy density for configurations of one and two magnetic vortices (Fig. \ref{fig:onetwo}), as a function of dot thickness. One can see from these curves that the configuration with more than one vortex, for ellipses with $2a=2560$ nm, corresponds to a minimum in energy only for thickness above about $50$ nm.

Finally, the effects of discretization, inherent in the methodology used here (see \cite{Jubert:2004}), are responsible for very small changes in the boundaries of our phase diagram for circular and elliptical nanodots, since on the one hand, the reduction of the dimension of the cell for the regions where the vortex corresponds to the minimum energy still results in this same configuration as minimum; on the other hand, where the vortex is not a minimum this is not altered by choosing a smaller cell size. These effects do not exist for perpendicular magnetization.

% --------------------------------------------------------------%
\section{Conclusions}
% --------------------------------------------------------------%

Using 3D numerical simulation we have obtained curves of energy per unit volume versus dimension for nanodots of elliptical and circular shape that illustrate how energy considerations explain the variability in magnetic arrangement found in these magnetic nanostructures.

We have shown in the present work the regions of geometrical parameters where soft magnetic dots of nanoscopic or mesoscopic dimensions, of circular and elliptical shapes, take up single domain and vortex magnetization configurations. For circular dots our results are in agreement with previous work performed using different methods. For elliptical dots there are some important differences, such as the appearance of a region where a magnetic structure with lateral vortices corresponds to an energy minimum.

Investigations that allow the mapping of these different magnetic configurations are useful in designing experiments to study the basic properties of these novel magnetic structures, or tailoring them for technological applications, such as magnetic random access memories.

% --------------------------------------------------------------%
\section*{Acknowledgments}
% --------------------------------------------------------------%
The authors would like to acknowledge the support of the Brazilian agencies CAPES, CNPq and FAPERJ, and comments on the manuscript made by L.C. Sampaio and J.P. Sinnecker.

\bibliographystyle{plain}
%\bibliography{arxiv}  % associado ao arquivo: 'bibliografia.bib'

\begin{thebibliography}{}

\end{thebibliography}


\begin{thebibliography}{17}
\expandafter\ifx\csname natexlab\endcsname\relax\def\natexlab#1{#1}\fi
\expandafter\ifx\csname bibnamefont\endcsname\relax
  \def\bibnamefont#1{#1}\fi
\expandafter\ifx\csname bibfnamefont\endcsname\relax
  \def\bibfnamefont#1{#1}\fi
\expandafter\ifx\csname citenamefont\endcsname\relax
  \def\citenamefont#1{#1}\fi
\expandafter\ifx\csname url\endcsname\relax
  \def\url#1{\texttt{#1}}\fi
\expandafter\ifx\csname urlprefix\endcsname\relax\def\urlprefix{URL }\fi
\providecommand{\bibinfo}[2]{#2}
\providecommand{\eprint}[2][]{\url{#2}}

\bibitem{Chien:2007}
\bibinfo{author}{\bibfnamefont{C.~L.} \bibnamefont{Chien}},
  \bibinfo{author}{\bibfnamefont{F.~Q.} \bibnamefont{Zhu}}, \bibnamefont{and}
  \bibinfo{author}{\bibfnamefont{J.-G.} \bibnamefont{Zhu}},
  \bibinfo{journal}{Physics Today} \textbf{\bibinfo{volume}{60}},
  \bibinfo{pages}{40} (\bibinfo{year}{2007}).

\bibitem{Shinjo:2000}
\bibinfo{author}{\bibfnamefont{T.}~\bibnamefont{Shinjo}},
  \bibinfo{author}{\bibfnamefont{T.}~\bibnamefont{Okuno}},
  \bibinfo{author}{\bibfnamefont{R.}~\bibnamefont{Hassdorf}},
  \bibinfo{author}{\bibfnamefont{K.}~\bibnamefont{Shigeto}}, \bibnamefont{and}
  \bibinfo{author}{\bibfnamefont{T.}~\bibnamefont{Ono}},
  \bibinfo{journal}{Science} \textbf{\bibinfo{volume}{289}},
  \bibinfo{pages}{930} (\bibinfo{year}{2000}).

\bibitem{Choe:2004}
\bibinfo{author}{\bibfnamefont{S.-B.} \bibnamefont{Choe}},
  \bibinfo{author}{\bibfnamefont{Y.}~\bibnamefont{Acremann}},
  \bibinfo{author}{\bibfnamefont{A.}~\bibnamefont{Scholl}},
  \bibinfo{author}{\bibfnamefont{A.}~\bibnamefont{Bauer}},
  \bibinfo{author}{\bibfnamefont{A.}~\bibnamefont{Doran}},
  \bibinfo{author}{\bibfnamefont{J.}~\bibnamefont{Stohr}}, \bibnamefont{and}
  \bibinfo{author}{\bibfnamefont{H.~A.} \bibnamefont{Padmore}},
  \bibinfo{journal}{Science} \textbf{\bibinfo{volume}{304}},
  \bibinfo{pages}{420} (\bibinfo{year}{2004}).

\bibitem{Hubert:1999}
\bibinfo{author}{\bibfnamefont{A.}~\bibnamefont{Hubert}} \bibnamefont{and}
  \bibinfo{author}{\bibfnamefont{R.}~\bibnamefont{Schäfer}},
  \emph{\bibinfo{title}{Magnetic {D}omains. {The} {A}nalysis of {M}agnetic
  {M}icrostructures}} (\bibinfo{publisher}{Springer},
  \bibinfo{address}{Berlin}, \bibinfo{year}{1999}).

\bibitem{Thomson:2008}
\bibinfo{author}{\bibfnamefont{T.}~\bibnamefont{Thomson}},
  \bibinfo{author}{\bibfnamefont{L.}~\bibnamefont{Abelman}}, \bibnamefont{and}
  \bibinfo{author}{\bibfnamefont{H.}~\bibnamefont{Groenland}}, in
  \emph{\bibinfo{booktitle}{Magnetic Nanostructures in Modern Technology}},
  edited by \bibinfo{editor}{\bibfnamefont{B.}~\bibnamefont{Azzerboni}},
  \bibinfo{editor}{\bibfnamefont{G.}~\bibnamefont{Asti}},
  \bibinfo{editor}{\bibfnamefont{L.}~\bibnamefont{Pareti}}, \bibnamefont{and}
  \bibinfo{editor}{\bibfnamefont{M.}~\bibnamefont{Ghidini}}
  (\bibinfo{publisher}{Springer}, \bibinfo{address}{Dordrecht},
  \bibinfo{year}{2008}), pp. \bibinfo{pages}{237--306}.

\bibitem{Kim:2008}
\bibinfo{author}{\bibfnamefont{S.}~\bibnamefont{Kim}},
  \bibinfo{author}{\bibfnamefont{K.}~\bibnamefont{Lee}},
  \bibinfo{author}{\bibfnamefont{Y.}~\bibnamefont{Yu}}, \bibnamefont{and}
  \bibinfo{author}{\bibfnamefont{Y.}~\bibnamefont{Choi}},
  \bibinfo{journal}{Appl. Phys. Lett.} \textbf{\bibinfo{volume}{92}},
  \bibinfo{pages}{022509} (\bibinfo{year}{2008}).

\bibitem{Bohlens:2008}
\bibinfo{author}{\bibfnamefont{S.}~\bibnamefont{Bohlens}},
  \bibinfo{author}{\bibfnamefont{B.}~\bibnamefont{Krüger}},
  \bibinfo{author}{\bibfnamefont{A.}~\bibnamefont{Drews}},
  \bibinfo{author}{\bibfnamefont{M.}~\bibnamefont{Bolte}},
  \bibinfo{author}{\bibfnamefont{G.}~\bibnamefont{Meier}}, \bibnamefont{and}
  \bibinfo{author}{\bibfnamefont{D.}~\bibnamefont{Pfannkuche}},
  \bibinfo{journal}{Appl. Phys. Lett.} \textbf{\bibinfo{volume}{93}},
  \bibinfo{pages}{142508} (\bibinfo{year}{2008}).

\bibitem{Ross:2002}
\bibinfo{author}{\bibfnamefont{C.~A.} \bibnamefont{Ross}},
  \bibinfo{author}{\bibfnamefont{M.}~\bibnamefont{Hwang}},
  \bibinfo{author}{\bibfnamefont{M.}~\bibnamefont{Shima}},
  \bibinfo{author}{\bibfnamefont{J.~Y.} \bibnamefont{Cheng}},
  \bibinfo{author}{\bibfnamefont{M.}~\bibnamefont{Farhoud}},
  \bibinfo{author}{\bibfnamefont{T.~A.} \bibnamefont{Savas}},
  \bibinfo{author}{\bibfnamefont{H.~I.} \bibnamefont{Smith}},
  \bibinfo{author}{\bibfnamefont{W.}~\bibnamefont{Schwarzacher}},
  \bibinfo{author}{\bibfnamefont{F.~M.} \bibnamefont{Ross}},
  \bibinfo{author}{\bibfnamefont{M.}~\bibnamefont{Redjdal}},
  \bibnamefont{et~al.}, \bibinfo{journal}{Phys. Rev. B}
  \textbf{\bibinfo{volume}{65}}, \bibinfo{pages}{144417}
  (\bibinfo{year}{2002}).

\bibitem{Metlov:2002}
\bibinfo{author}{\bibfnamefont{K.~L.} \bibnamefont{Metlov}} \bibnamefont{and}
  \bibinfo{author}{\bibfnamefont{K.~Y.} \bibnamefont{Guslienko}},
  \bibinfo{journal}{J. Magn. Magn. Mat.} \textbf{\bibinfo{volume}{242-245}},
  \bibinfo{pages}{1015} (\bibinfo{year}{2002}).

\bibitem{Scholz:2003}
\bibinfo{author}{\bibfnamefont{W.}~\bibnamefont{Scholz}},
  \bibinfo{author}{\bibfnamefont{K.}~\bibnamefont{Guslienko}},
  \bibinfo{author}{\bibfnamefont{V.}~\bibnamefont{Novosad}},
  \bibinfo{author}{\bibfnamefont{D.}~\bibnamefont{Suess}},
  \bibinfo{author}{\bibfnamefont{T.}~\bibnamefont{Schrefl}},
  \bibinfo{author}{\bibfnamefont{R.}~\bibnamefont{Chantrell}},
  \bibnamefont{and} \bibinfo{author}{\bibfnamefont{J.}~\bibnamefont{Fidler}},
  \bibinfo{journal}{J. Magn. Magn. Mat.} \textbf{\bibinfo{volume}{266}},
  \bibinfo{pages}{155} (\bibinfo{year}{2003}).

\bibitem{Zhang:2008}
\bibinfo{author}{\bibfnamefont{W.}~\bibnamefont{Zhang}},
  \bibinfo{author}{\bibfnamefont{R.}~\bibnamefont{Singh}},
  \bibinfo{author}{\bibfnamefont{N.}~\bibnamefont{Bray-Ali}}, \bibnamefont{and}
  \bibinfo{author}{\bibfnamefont{S.}~\bibnamefont{Haas}},
  \bibinfo{journal}{Phys. Rev. B} \textbf{\bibinfo{volume}{77}},
  \bibinfo{eid}{144428} (pages~\bibinfo{numpages}{8}) (\bibinfo{year}{2008}),
  \urlprefix\url{http://link.aps.org/abstract/PRB/v77/e144428}.

\bibitem{Buchanan:2005}
\bibinfo{author}{\bibfnamefont{K.~S.} \bibnamefont{Buchanan}},
  \bibinfo{author}{\bibfnamefont{P.~E.} \bibnamefont{Roy}},
  \bibinfo{author}{\bibfnamefont{M.}~\bibnamefont{Grimsditch}},
  \bibinfo{author}{\bibfnamefont{F.~Y.} \bibnamefont{Fradin}},
  \bibinfo{author}{\bibfnamefont{K.~Y.} \bibnamefont{Guslienko}},
  \bibinfo{author}{\bibfnamefont{S.~D.} \bibnamefont{Bader}}, \bibnamefont{and}
  \bibinfo{author}{\bibfnamefont{V.}~\bibnamefont{Novosad}},
  \bibinfo{journal}{Nature Phys.} \textbf{\bibinfo{volume}{1}},
  \bibinfo{pages}{172} (\bibinfo{year}{2005}).

\bibitem{Lai:2007}
\bibinfo{author}{\bibfnamefont{M.-F.} \bibnamefont{Lai}},
  \bibinfo{author}{\bibfnamefont{Z.-H.} \bibnamefont{Wei}},
  \bibinfo{author}{\bibfnamefont{J.~C.} \bibnamefont{Wu}},
  \bibinfo{author}{\bibfnamefont{W.~Z.} \bibnamefont{Shieh}},
  \bibinfo{author}{\bibfnamefont{C.~R.} \bibnamefont{Chang}}, \bibnamefont{and}
  \bibinfo{author}{\bibfnamefont{J.}~\bibnamefont{Guo}}, \bibinfo{journal}{J.
  Appl. Phys.} \textbf{\bibinfo{volume}{101}}, \bibinfo{pages}{09N111}
  (\bibinfo{year}{2007}).

\bibitem{Usov:2001}
\bibinfo{author}{\bibfnamefont{N.~A.} \bibnamefont{Usov}},
  \bibinfo{author}{\bibfnamefont{C.-R.} \bibnamefont{Chang}}, \bibnamefont{and}
  \bibinfo{author}{\bibfnamefont{Z.-H.} \bibnamefont{Wei}},
  \bibinfo{journal}{J. Appl. Phys.} \textbf{\bibinfo{volume}{89}},
  \bibinfo{pages}{7591} (\bibinfo{year}{2001}),
  \urlprefix\url{http://link.aip.org/link/?JAP/89/7591/1}.

\bibitem{Ha:2003}
\bibinfo{author}{\bibfnamefont{J.~K.} \bibnamefont{Ha}},
  \bibinfo{author}{\bibfnamefont{R.}~\bibnamefont{Hertel}}, \bibnamefont{and}
  \bibinfo{author}{\bibfnamefont{J.}~\bibnamefont{Kirschner}},
  \bibinfo{journal}{Phys. Rev. B} \textbf{\bibinfo{volume}{67}},
  \bibinfo{pages}{224432} (\bibinfo{year}{2003}).

\bibitem{Jubert:2004}
\bibinfo{author}{\bibfnamefont{P.~O.}~\bibnamefont{Jubert}} \bibnamefont{and}
  \bibinfo{author}{\bibfnamefont{R.}~\bibnamefont{Allenspach}},
  \bibinfo{journal}{Phys. Rev. B} \textbf{\bibinfo{volume}{70}},
  \bibinfo{pages}{144402} (\bibinfo{year}{2004}).

\bibitem{Cowburn:1999}
\bibinfo{author}{\bibfnamefont{R. P.} \bibnamefont{ Cowburn}},
  \bibinfo{author}{\bibfnamefont{D. K}~\bibnamefont{Koltsov}},
  \bibinfo{author}{\bibfnamefont{A. O.}~\bibnamefont{Adeyeye}},
  \bibinfo{author}{\bibfnamefont{M. E.}~\bibnamefont{Welland}},
  \bibinfo{author}{\bibfnamefont{D. M}~\bibnamefont{Tricke}},
    \bibinfo{journal}{Phys. Rev. Lett.} \textbf{\bibinfo{volume}{83}},
  \bibinfo{pages}{1042-10450} (\bibinfo{year}{1999}).

\bibitem{oommf} \url{http://math.nist.gov/oommf/}.

\end{thebibliography}

\end{document}